\documentclass[12pt,manuscript]{aastex}

\shorttitle{Parameters of Procyon}
\shortauthors{Liebert, Fontaine et al.}

\begin{document}

\title{The Age and Stellar Parameters of the Procyon Binary System}

\author{James Liebert\altaffilmark{1}, Gilles Fontaine\altaffilmark{2}, 
Patrick A. Young\altaffilmark{3}, Kurtis A. Williams\altaffilmark{4}, 
and David Arnett\altaffilmark{1} }

\slugcomment{\bf Accepted for ApJ, 15 Mar 2013} 

\altaffiltext{1}{Department of Astronomy and Steward Observatory,
University of Arizona, Tucson, AZ 85721, jamesliebert@gmail.com,
darnett@as.arizona.edu }

\altaffiltext{2}{D\'epartement de Physique, Universit\'e de
Montr\'eal, C.P. 6128, Succ. Centre-Ville, Montr\'eal, Qu\'ebec, 
Canada H3C 3J7, fontaine@astro.umontreal.ca }

\altaffiltext{3}{School of Earth and Space Exploration, 
Arizona State University, Tempe AZ 85287, pyoung.3@asu.edu }

\altaffiltext{4}{Texas A\&M University, Commerce TX 75429,
Kurtis.Williams@tamuc.edu }

\begin{abstract}

The Procyon~AB binary system (orbital period 40.838 years, a
newly-refined determination), is near and bright enough that the
component radii, effective temperatures and luminosities are very well
determined, although more than one possible solution to the masses has
limited the claimed accuracy.  Preliminary mass determinations for
each component are available from HST imaging, supported by
ground-based astrometry and an excellent {\it Hipparcos} parallax; we
use these for our preferred solution for the binary system.  Other
values for the masses are also considered.  We have employed the TYCHO
stellar evolution code to match the radius and luminosity of the
F5~IV-V primary star to determine the system's most likely age as
1.87$\pm$0.13~Gyr.  Since prior studies of Procyon~A found its
abundance indistinguishable from solar, the solar composition of
Asplund, Grevesse \& Sauval ($Z$=0.014) is assumed for the HR Diagram
fitting.  An unsuccessful attempt to fit using the older solar
abundance scale of Grevesse \& Sauval ($Z$=0.019) is also reported.
For Procyon~B, eleven new sequences for the cooling of non-DA white
dwarfs have been calculated, to investigate the dependences of the
cooling age on (1) the mass, (2) the core composition, (3) the helium
layer mass, and (4) heavy-element opacities in the helium envelope.
Our calculations indicate a cooling age of 1.19$\pm$0.11 Gyr, which
implies that the progenitor mass of Procyon~B was
2.59$_{-0.26}^{+0.44}$ $M_{\odot}$. In a plot of initial vs final mass
of white dwarfs in astrometric binaries or star clusters (all with age
determinations), the Procyon~B final mass lies several $\sigma$ below
a straight line fit.

\end{abstract}


\section{Introduction}

Procyon~A is the 14th nearest star and stellar system to the Sun,
according to Henry (2011), at a distance of 3.51 parsecs or 11.46
light years -- see van Leeuwen (2007).  Its spectral type is F5~IV-V.
Bessel (1844) discovered that the (large) proper motion of Procyon on
the sky was perturbed by what he recognized to be an unseen companion.
Procyon~B was first detected visually at the end of the 19th century
by Schaeberle (1896). The first recognition that the companion was
like those in the Sirius and 40~Eri systems, later to be called white
dwarfs, may be the work of the Irish amateur astronomer John Ellard
Gore.  However, he did not properly publish this result in the
astronomical literature -- see the discussions of his work by
FitzGerald (1966) and Holberg (2009). 

The first orbital elements were determined by Auwers (1862), who
showed that the orbital period would be about 40 years.  We now know
that the more exact period is 40.838 years (H. E. Bond \&
G.H. Schaefer, 2012 private communication). Solar-like oscillations in
Procyon A were first reported by Brown et al. (1991), and several
authors have since investigated the structure and evolution of that star
through a seismic approach (see, e.g., Dogan et al. 2010 and references
therein). We come back to the asteroseismological properties of Procyon
A in \S~5. 

Like Sirius~A studied in Liebert et al. (2005, hereafter Paper~I),
Procyon~A has an extremely accurate interferometric radius
measurement, and an extremely accurate trigonometric parallax; the
luminosity is also very accurate.  Sirius~A is certainly a main
sequence star, while Procyon~A could be entering or about to enter the
subgiant branch.  In contast to the Sirius system, however, the
astrometric mass solution for the Procyon system has been a
controversial issue, as we shall discuss in \S~2.

In comparison with the white dwarf Sirius~B, Procyon~B is
very different from other (mostly DA) white dwarfs previously used for
the problem of the star's initial-to-final mass relation (IFMR).
Mainly it has a helium-dominated atmosphere polluted by heavy elements
(DQZ spectral type), a lower mass, and a cooler temperature.  The mass,
radius and $T_{\rm  eff}$ determined for the B component will be
discussed in \S~3. 

In \S~4 we address the atmospheric abundances of Procyon~A and the
interior abundances to be used for modelling in \S~5.  Here we employ
the TYCHO stellar evolution code (Young \& Arnett 2005) to fit the
position of Procyon~A in an HR diagram, and determine the age.

In \S~6 eleven calculations of the cooling time of Procyon~B are
presented, for different assumptions about the mass, the helium
envelope mass, and its composition.  The cooling time is the length of
time since the degenerate entered the white dwarf sequence; this has
been used for the coolest white dwarfs to estimate ages of the
Galactic disk (Liebert, Dahn, \& Monet 1988) and halo (Hansen et
al. 2007 -- using the globular cluster NGC~6397). The difference
between the systemic age and the white dwarf cooling time yields the
best theoretical estimate of the initial mass of the white dwarf.
Moreover, as discussed in \S~6, Procyon~B presents a complicated case
for the calculation of the cooling age. Further remarks pertinent to
this more challenging physical situation are discussed in that
section.

This paper adds a valuable new object with accurately determined
stellar parameters (masses, radii, luminosities) to the
initial-to-final mass relation (IFMR) for progenitors and white
dwarfs.  The vast majority of other data points have white dwarfs with
hydrogen-rich atmospheres, while Procyon B has a complicated
helium-rich envelope and atmosphere.  It is possible that the IFMR for
helium-rich white dwarfs may differ from the hydrogen-rich
distribution.  We show in \S~7 how Procyon~B compares with Sirius~B
and white dwarfs in star clusters for which the IFMR of low to
intermediate mass stars has been determined.

\section{The Physical Parameters of Procyon A}

As noted in the Introduction, a precise radius measurement is
available for Procyon~A, from measurements of the angular diameter
with the ESO Very Large Telescope Interferometer.  The radius is $R$ =
2.031$\pm$0.013$R_{\odot}$~or $\pm$0.65\% (Aufdenberg, Ludwig, \&
Kervella 2005).  This radius value is based on the authors ``best
estimate'' angular diameter of 5.404$\pm$0.031 mas at 2.2$\mu$.  Note
that the authors constructed model atmospheres for Procyon in three
ways: (1) stand-alone one-dimensional (1D) PHOENIX structures,
radiation fields, and spectra (Hauschildt et al. 1999); (2) CO$^5$BOLD
3D structures temporarily and spatially averaged to 1D, then read by
PHOENIX for computation of the corresponding radiation fields (Freytag
et al. 2002); and (3) stand-alone 1D ATLAS~12 (Kurucz 1992)
structures, radiation fields and spectra.  Note that the quantitative
assessment of the limb darkening, on which the radius value is
determined, depends on the model atmospheres analysis.  The authors
discuss in detail their assessment of these different models, leading
to their ``best estimate''.

These authors also cite the earlier measurement on Mt. Wilson,
California using the Mark III Interferometer at 500 and 800~nm
(Mozurkewich et al 1991), whose limb-darkened value of the angular
diameter (5.51$\pm$0.05 mas) agrees marginally within the errors.
Their new measurement compares to an older, slightly less accurate,
but consistent value of 2.048$\pm$0.025$R_{\odot}$~ by the same group
(Kervella et al. 2004), which corresponds to an adopted angular
diameter of 5.448$\pm$0.053 mas.  Casagrande et al. (2010) got a
lower, less accurate, but consistent 5.326$\pm$0.068 mas.  For the
radius ($R_A$) of the primary star in this paper, we adopt the
Aufbenberg et al. (2005) best value.  Note that the Kervella et
al. and Aufdenberg et al. results differ by less than 1$\%$.  Adopting
either one results in an essentially identical fit.

The {\it Hipparcos} satellite mission provided the precise measurement
of the trigonometric parallax of 0.28593$\pm$0.00088" (or $\pm$0.3\%).
Using the bolometric energy distribution and the effective temperature
determination (6,543$\pm$84~K), the luminosity of Procyon~A is
determined in Aufdenberg et al. (2005) to be log $L$/$L_{\odot}$ =
+0.83$\pm$0.04.  Note that Chiavassa et al. estimate $T_{\rm eff}$ =
6591~K or 6556~K depending on how the bolometric flux is calculated,
within the error bars of Aufdenberg et al. (2005).  The former value
differs the most from Aufdenberg et al.'s preferred 6543~K, resulting
in log $L$/$L_{\odot}$ = +0.843 (with one more significant digit than
is warranted).  The difference between these two values hardly matters
in Figure 1.

Two decades ago the mass of the ``A'' component was believed to be
near 1.75~$M_{\odot}$ (Irwin et al. 1992).  However, main sequence
evolutionary models at that mass were too luminous by a factor of two
compared with the observed luminosity (Guenter \& Demarque 1993).
Long-term imaging by HST, combined with the interferometric radius and
trigonometric parallax measurements cited above led to a great
improvement in the stellar parameters.  First, Girard et al.(2000)
produced an improved astrometric solution based on analysis of 250
photographic plates spanning 83 years, augmented with early Hubble
data.  They found M$_A$ = 1.497$\pm$0.037 $M_{\odot}$, almost exactly
the 1.50 $M_{\odot}$ predicted by Guenther \& Demarque (1993).

The problem is that HST imaging has only been possible for only about
half of the orbital period.  The combining of space-based imaging and
the much longer series of ground-based observations has often been
fraught with difficulties.  Apparent systematic offsets often 
occur in the respective measurements of the separations and position
angles.  Girard et al. (2000) found that, if they restricted the
astrometric analysis to the excellent images obtained with the WFPC2
camera -- i.e., with no ground-based data -- a substantially lower mass
(1.465 $M_{\odot}$) for Procyon A was the result.  At that time HST
imagery covered only about a decade.  

Allende Prieto et al. (2002) revisited the Procyon A solution using
the {\it Hipparcos} parallax and the stellar angular diameter by
Mozurkevich et al. (1991) discussed previously.  They derived the
primary's mass to be a much lower 1.42$\pm$0.06 $M_{\odot}$.  This is
consistent with the ground-based Gatewood and Han (2006) value of the
primary's mass of 1.43$\pm$0.034 $M_{\odot}$, determined from their
Multichannel Astrometric Photometer {\bf MAP} camera, built at the
University of Pittsburg (Gatewood 1987).  The {\bf MAP} parallax is
286$\pm$0.95 mas, which agrees well with the {\it Hipparcos} value.

Schaefer et al. (2006) reported a preliminary solution resulting from
a Space Telescope Science Institute-led program to extend the imaging
program on the Procyon binary orbit.  Their result was in excellent
agreement with the Girard et al. (2000) analysis including
ground-based data.  Several more years of HST data since 2006 now
result in the following masses: 1.499$\pm$0.031 $M_{\odot}$ for
Procyon~A and 0.553$\pm$0.022 $M_{\odot}$ for Procyon~B (G. H. Schaefer
and H. E. Bond, 2012 private communication).  This solution uses
available HST data from 1995 through 2012 plus, appropriately
weighted, the archival ground-based data.

It is not difficult to see how the total mass of the system is
specified from accurate HST astrometry.  Using Newton's
gravitational formula for Kepler's third law, ($M_A$ +$M_B$) = 
$a^3$~/~$\pi^3$~$P^2$, with the semimajor axis ($a$) and
period ($P$) well known, the sum of the masses is fixed at
2.052 $M_{\odot}$.  Ground-based observations of the wobble of the A
component about the center-of-mass on the sky then fix $a_A$,
$a_A$~+~$a_B$~=~$a$, and $M_A$/$M_B$~=~$a_B$/$a_A$.  Thus the solution
provides separate values for $M_A$ and $M_B$.

The mass determinations remain a matter of dispute in the literature.
We begin with 1.499 $M_{\odot}$ as the HST-based, preferred
mass of the primary for fitting in \S~5.  We also consider 1.42
$M_{\odot}$ as a lower bound to the mass of the primary.  Note that,
in their recent analysis, Chiavassa et al. (2012) give arguments in
favor of the 1.43 $M_{\odot}$ Gatewood and Han value.

\section{The Physical Parameters of Procyon B}

Procyon B was first detected visually at the end of the 19th century 
by Schaeberle (1896), and was historically one of the first white 
dwarfs to be discovered. Even early estimates of the mass ($M_B$) of the
white dwarf component, such as that by Strand (1951), showed that it is
much closer to the mean value of field white dwarfs than is Sirius~B.
The updated Schaefer et al. HST solution gives 0.553$\pm$0.022
$M_{\odot}$ for the mass of the secondary, and we consider this the
benchmark value for $M_B$.  To be sure this is much smaller than
previous estimates for the secondary, such as 0.622 $M_{\odot}$ (Irwin
et al. 1992) and 0.602$\pm$0.015 $M_{\odot}$ (Girard et al. 2000).
Values in between these have been published.  The Kervella et al. (2004)
lower mass solution of 1.42 $M_{\odot}$ for the primary led to $M_B$ =
0.575$\pm$0.017 $M_{\odot}$ for B.  The Gatewood and Han (2006) primary
mass of 1.43$\pm$0.034 $M_{\odot}$ resulted in 0.58$\pm$0.014
$M_{\odot}$ for the secondary.  To illustrate the dependence on mass,
the cooling time for a mass of 0.602 $M_{\odot}$ (Girard et al. 2000) is
also discussed in \S~6. 

As mentioned in the Introduction, the nature of the ``B'' white dwarf
poses additional problems.  The atmosphere is helium-rich, with trace
abundances of carbon, calcium, magnesium and iron, but no detection of
hydrogen.  This is an unusually diverse abundance pattern for a white
dwarf atmosphere, and complicated greatly the determination of the
atmospheric parameters.  Not knowing of the complication, in the
absence of an ultraviolet spectrum, Provencal et al. (1997) used pure
hydrogen and helium atmospheres to model the spectral energy
distribution.  Preferring the helium fit due to its optical ``DC''
spectrum, they found $T_{\rm eff}$ = 8,688~K.  This implied a stellar
radius of 0.0096$\pm$0.0005 $R_{\odot}$.  Given the  mass, this small 
radius implied that the interior composition of the white dwarf could
not be carbon-oxygen, but rather something very dense like iron
(Shipman \& Provencal 1999).  This conclusion was unsettling to
stellar theorists.
      
Once the complicated ultraviolet spectrum was revealed with the Space
Telescope Imaging Spectrograph on the Hubble Space Telescope,
Provencal et al. (2002) got themselves out of this ``iron box''.
Atmospheres with realistic trace adundances, obtained from fitting the
spectrum and the 1,800--10,000\AA\ spectral energy distribution were
employed to obtain $T_{\rm eff}$ = 7,740$\pm$50~K.  They now estimated
the radius to be 0.01234$\pm$0.00032 $R_{\odot}$, consistent with a
carbon-oxygen interior.  Note that Procyon~B is the only white dwarf in
a binary system or star cluster which does not have an atmosphere
amenable to a mass determination from either the fitting of hydrogen or
He~I lines. It is also the coolest such white dwarf.

\section{The Atmospheric and Interior Abundances of Procyon A}

There is a long literature trail indicating that the Procyon~A
atmospheric abundance cannot be distinguished from solar.  We cite
five studies: Heiter \& Luck (2003), and Luck \& Heiter (2005), as
part of a study of all bright, nearby stars in the Northern
Hemisphere, analyzed Procyon~A using high resolution spectra.  They
found an iron abundance [Fe/H] = $-$0.04$\pm$0.06.  Valenti \& Fischer
(2005), as part of a systematic study of nearby stars using echelle
spectra, obtained [Fe/H] = $-$0.05$\pm$0.03.  In Allende Prieto et
al.'s (2004) study of stars more luminous than M$_V$ = +6.5 within
14.5~pc of the Sun, a value of +0.08 was determined for
Procyon~A. Finally, in the most comprehensive Geneva-Copenhagen study
of solar neighborhood stars (Nordstr\"om et al. 2004), a value of
[Fe/H] = +0.05 was reported.  Both of the last two author groups
discuss the possible systematic error due to the coupling of the
[Fe/H] determination to the error in the log $g$ value.  Note that
Procyon~A is cool enough that peculiar diffusion / gravitational
support mechanisms should not distort the surface abundances from the
interior abundance.  There could, nevertheless, be a small diffusion
of heavy elements out of the thin convective envelope.

However, recently there has been controversy about what the solar
abundance actually is.  Inevitably, the abundance parameters to be run
for the stellar evolution models of Procyon~A, to be described next in
\S~5, must therefore be linked to this controversy, so a few comments
are in order.

The newer solar abundance scale determined from 3D, non-LTE calculations 
of the solar atmosphere (Asplund et al. 2004, 2005, 2009 -- the
latter hereafter A09), and now Lodders 2010) have significantly lower
abundances of oxygen, carbon and nitrogen than previous estimates
(Grevesse \& Sauval 1998).  The abundance of neon is a particular
problem, since it is impossible to measure this element in the Sun (or
Procyon~A), but is likely to be comparable numerically to the three
elements mentioned previously.  The overall heavy element abundance
parameter for the Sun decreases to a primordial value of $Z$=0.0153.

The Asplund A09 scale has been challenged because it breaks the
accordance of the ``standard solar'' interior model with
helioseismological measurements (Bahcall et al. 2005).  A solar model
with these abundances incorrectly predicts the depth of the convection
zone, the depth profiles of sound speed and density, and the helium
abundance (Basu \& Antia 2004).  As a check on the magnitude of likely
systematic errors due to uncertainty in the actual solar (and
Procyon~A) abundance, we retain consideration of the older Grevesse \&
Sauval (1998) solar abundance which corresponds to $Z$=0.019.

\section {The Fitting of Procyon~A in a Radius-Luminosity ``HR'' Diagram}

The TYCHO code -- http://chandra.as.arizona.edu/$\sim$dave/tycho-intro.html
-- incorporates the most current microphysics for opacities,
equation-of-state, and nuclear reactions.  In particular, the code has
been upgraded with an algorithm based on a physical analysis of 3D
hydrodynamic simulations of convection, not an
astronomically-calibrated version of the mixing length theory.  It
includes non-locality and time dependence of flow, dynamical
acceleration, turbulent dissipation, Kolmogorov heating, compositional
effects and dynamically-defined boundary conditions (instead of
parameterized overshooting schemes), all in a single, self-consistent
formulation.  The code is regularly tested against observations of
double-lined eclipsing binaries and cluster isochrones to ensure
consistency and accuracy.  It produces superior fits to observational
test cases without adjustment of parameters. 

These changes make possible fits to nuclear-burning stars on the HR
Diagram, especially those on or near the main sequence (Arnett,
Meakin, \& Young 2009, 2010).  Robust, consistent ages for
(nondegenerate) binary stars with well-determined luminosities,
masses, radii and temperatures have been determined by Young et
al. (2001), Young \& Arnett (2005).

In additon to the mass, we have to input the abundance parameters $Y$
and $Z$, where $Z$ scales by the solar ratios of individual elements.
Since we cannot regard the solar atmospheric abundance issue as fully
resolved, we therefore have computed sequences using the Lodders
(2010) value of $Z$ = 0.0153 and also evaluated the older solar
abundance of Grevesse and Sauval (1988) which imply $Z$ = 0.019.

In Figure~1 we show evolutionary tracks for $Z$ = 0.0153 of (1)
1.499 $M_{\odot}$ (in black), the best value from preliminary HST
astrometry (\S~2), fitting the correct luminosity and radius at age
1.74~Gyr; (2) 1.48 $M_{\odot}$ (red) at 1.87~Gyr, the best fit to the
luminosity (to three significant digits); (3) 1.463 $M_{\odot}$ (green)
at 1.99~Gyr; and (4) the lower 1.42 $M_{\odot}$ (blue) at 2.51~Gyr
favored by largely ground-based studies cited in \S~2.  The highest
three masses fit within 1$\sigma$ late on the main sequence.  For
1.42 $M_{\odot}$ the fit misses by more than 2$\sigma$ late on the main
sequence.  Note that best-fit ages include the pre-MS evolution,
without pre-MS accretion, but only the tracks from the beginning of
the main sequence are shown here.  We also tried a fit with the
Grevesse \& Sauval (1998) $Z$ = 0.019 abundance at 1.48 $M_{\odot}$, but
it is 1.45$\sigma$ underluminous when it reaches the correct radius on
the main sequence at age 1.96~Gyr.  Given the evidence from A09 and 
Lodders (2010) that this solar metallicity value is too large, 
this marginal attempt at a fit is not shown in the figure.

We therefore believe that 1.87$\pm$0.13~Gyr from the 1.48 $M_{\odot}$
fit is the best estimate of the age of the binary system, with the
uncertainty given by the bracketing age values for the 1.499 and
1.463 $M_{\odot}$ fits.  Note that Provencal et al. (2002) estimated
the nuclear lifetime of 1.3~Gyr for component B, estimating
$\sim$2.1 $M_{\odot}$ as its progenitor mass.  These authors use the
cooling age from models of Wood (1995) for pure carbon cores -- the
only calculations available then.  We now explore a range of
alternative compositions for the white dwarf in \S~6. Since our
best-fit mass using the TYCHO models lies well within the error bars
of the preliminary HST astrometric mass of 1.499 $M_{\odot}$, we adopt
1.48 $M_{\odot}$ as the preferred or ``benchmark'' mass of Procyon~A.

Note that the fitting of a star with mass near 1.5 $M_{\odot}$ at this
stage of evolution presents uncertainties due to having both a small
convective core and a small convective envelope.  Due to the former,
the main sequence lifetime is affected by the uncertainty in the
amount of hydrogen fuel available for reactions in the core.  Note
that remaining uncertainties in the treatment of mixing beyond the
convective core boundary, in the primordial solar abundance mix, or in
any of the other model physics probably add larger systematic errors,
by amounts which are difficult to quantify.  The convective envelope
leads to implied small differences between the interior and
atmospheric $Z$ abundances, as noted earlier.

For Kervella et al.'s (2004) similar analysis, the fit to the HRD
parameters also occurs in a similar way.  If 1.50 $M_{\odot}$ is
assumed, their fit is late on the main sequence at the much younger
age of 1.3~Gyr.  However, these authors assumed {\it no overshooting}
of the convective core, i.e., less fuel, vs our calculations.  For
their preferred mass of 1.42 $M_{\odot}$, the corresponding fit is
early in the H-shell burning phase at ages of 2.31--2.71~Gyr.  This
age range is close to what we achieve at 1.42$M_{\odot}$ for which 
we determine an age of 2.51~Gyr (stated three paragraphs earlier),  
plotted as the blue track in Figure 1.

As indicated in the Introduction, stochastically excited $p$-modes have
been observed in Procyon A since first reported by Brown et al. (1991),
and several authors have attempted to exploit the seismic potential of
that star using different data sets and models (e.g., Guenther \&
Demarque 1993; Barban et al. 1999; Chaboyer et al. 1999; Di Mauro \&
Christensen-Dalsgaard 2001; Eggenberger et al. 2005; Provost et
al. 2006; Bonanno et al. 2007). However, their investigations have been
limited by the accuracy of the detected pulsation frequencies. More
recently, better data became available (Bedding et al. 2010) and have
been modelled in a preliminary way by Dogan et al. (2010). We note that
Dogan et al. (2010) report two possible seismic model for Procyon A, one
of which characterized by a total mass of 1.50 $M_{\odot}$ and an age of
1.83 Gyr (no uncertainties provided), which is remarkably close to our
solution. We find this result most encouraging.

\section{The cooling age and progenitor mass of Procyon~B}

\subsection{White dwarf evolutionary sequences}

The importance of the cooling age of a white dwarf was introduced in
the Introduction.  For Procyon~B we first calculate the cooling time
for a non-DA white dwarf of $M_B$ = 0.553$\pm$0.022 $M_{\odot}$, the
benchmark mass, but also evaluate the earlier alternative of $M_B$ =
0.602 $M_{\odot}$ from Kervella et al. (2000).  The fitted $T_{\rm
  eff}$ is 7,740$\pm$50~K, as discussed  in \S~3.  These error bars
are the published, internal errors of the fit to the spectral energy
distribution and absorption features, but are uncomfortably less than
1\% of the value itself.  For this object with a complicated DQZ
spectrum, we feel more comfortable assuming quadruple the formal
$T_{\rm eff}$ error or $\pm$200~K.

There have unfortunately been a shortage of available evolutionary
sequences for white dwarfs lacking thick hydrogen envelopes.  Most
cluster white dwarf studies in the last few decades have used the
evolutionary models of Wood (1992, 1995), but always with an outer
hydrogen layer.  These calculations do include one sequence with no
hydrogen, a ``thin'' helium envelope of 10$^{-4}$, atop a pure carbon
core but with solar abundances of elements heavier than helium (Wood
2012, private communication).  Hansen (1999) showed that pure helium
envelope cooling models for 0.6 $M_{\odot}$ by himself (for C, O and
C/O cores) and by Salaris et al. (1997) cool considerably faster than
the Wood sequence (see Hansen's Fig.~7).  Note that if a trace of
heavy elements were assumed, the envelope opacity would be
considerably increased.  Hence, there has been some confusion as to
what the cooling rates of white dwarfs lacking hydrogen layers should
be.  This issue is addressed in \S~6.2.

For a cool, non-DA white dwarf with a simple, featureless spectrum,
one could argue that the use of a pure-He envelope sequence such as
from Hansen (1999) would yield the more robust cooling time.  However,
the spectrum of Procyon~B shows the effects of (1) dredge-up of carbon
from the diffusion tail of the core, with an atmospheric abundance
log~[C/He]~= $-$5.5$\pm$0.2 (Provencal et al. 2002); and (2) accretion
of heavy elements -- log~Mg/He = $-$10.4, log~Ca/He = $-$12, log~Fe/He =
$-$10.7 with log~H/He $<$ $-$4.  Provencal et al. speculate that the
secondary could be accreting a stellar wind from the primary at a rate of
about 2$\times$~10$^{-19}M_{\odot}$~yr$^{-1}$.  Thus, a calculation
including additional opacity in the envelope layer may be more
appropriate.  Moreover, systematic errors need to be considered also
due to the uncertain mix of carbon and oxygen in the core, and the
thickness of the helium envelope.  For this purpose one of us (GF) has
calculated a new set of cooling tracks to estimate these errors and
parameter dependences, and hopefully thus bound the possible cooling
time of this white dwarf.

Salaris et al. (1997) explored the dependence of cooling rates on the
internal chemical distribution of carbon and oxygen in the core.  For
their best choices of the combined effect of convective mixing and the
$^{12}$C($\alpha$,$\gamma$)$^{16}$O reaction rate, carbon-oxygen
profiles showing an enhancement of oxygen in the central regions are
obtained for all white dwarf masses.  Mass fractions of $^{16}$O as
high as 0.8 are found near 0.6 $M_{\odot}$.  This fraction declines
with increasing core mass because the above reaction rate is highest
at the lower core temperatures characteristic of lower masses.  Thus,
a core composition dominated by oxygen, or at least a mixture of
carbon and oxygen, is more likely than one of pure carbon for
Procyon~B.  Sequences have been calculated for pure oxygen, pure
carbon, and for mixtures of carbon and oxygen according to the 
mass-dependent and depth-dependent results of Salaris et al. (1997).
We pick the last of these three types of sequences as the benchmark core
composition. 

The cooling time also depends on the thickness of the helium layer.
The helium-rich atmospheric composition indicates that some of the
helium layer mass could have been lost in a late helium-shell flash
that disposed of all of the hydrogen (Iben 1984).  Values in the
literature that have been considered include masses between 10$^{-2}$ and
10$^{-4}M_{\odot}$.  The atmospheric carbon abundance is a strong
constraint, since this results from dredgeup of the diffusion tail of
carbon from the core.  The study (Fontaine \& Brassard 2005) fitting
dredgeup models to match the spectra of Procyon~B and a sequence of
field DQ white dwarfs by Dufour, Bergeron, \& Fontaine (2005) shows
that the atmospheric carbon abundance of Procyon~B favors a layer mass
thickness of log~$M_{He}/M_{\odot}\sim$$-$2.5$\pm$0.5 (see their Fig.~12).
We therefore consider this value as the benchmark, with sequences also
calculated for a reasonable range of $-$2.0 and $-$3.0.

Finally, it is very difficult to estimate the opacities due to
elements heavier than helium in the partially-degenerate helium layer.
This will depend on how deeply the heavy elements have diffused into
the envelope below the convection zone.  In principle the carbon
abundance profile can be calculated for the helium envelope from
diffusion theory, with the atmospheric abundance as a surface
boundary.  To consider this dependence in an approximate way,
sequences with $Z$=0.001 and $Z$=0 have been calculated, with the former
as the benchmark while the latter is contradicted by the observation
of carbon and heavier elements in the spectrum.
 
In Table 1, the cooling ages for the parameters of Procyon~B are
listed for 8 sequences at the favored mass of 0.553 $M_{\odot}$, two
for values at $\pm$0.022$M_{\odot}$ , and one for the alternative
0.602 $M_{\odot}$ (Girard et al. 2004).  Also listed for each sequence
are the core composition (C, O, or a more realistic Salaris C/O mix),
the helium layer mass ($-$2.5$\pm$0.5 as discussed earlier), and the
heavy element abundance $Z$=0.001 or 0.  All assume zero hydrogen
abundance.

The benchmark cooling age of Procyon~B is given by the model for
0.553 $M_{\odot}$, with C/O core, log~M(He)/M = $-$2.5, and $Z$ = 0.001 in
the envelope (benchmark parameters). The cooling value is 1.187~Gyr,
with a formal, internal error of $\pm$0.104~Gyr due to uncertainties in
the assumed effective temperature and the assumed mass. According to
Table 1, a range of $\pm$200 K about the estimated effective temperature
of $T_{\rm eff}$ = 7740 K translates into an age uncertainty of 
$\pm$0.085 Gyr, while a range of $\pm$0.022 $M_{\odot}$ about the
benchmark mass of 0.533 $M_{\odot}$ translates into an age interval of 
$\pm$0.061 Gyr. Added in quadrature, these uncertainties lead to
$\pm$0.104 Gyr. In addition, two sources of systematic error
should be considered here.  The age difference between a mixed (C/O)
and a pure oxygen core is relatively small at 0.010~Gyr. For its part,
uncertainty due to variations in the helium layer mass is
$\pm$0.043~Gyr. Added again in quadrature with the above internal error,
these systematic effects lead to a total uncertainty of $\pm$0.113 Gyr
on the age. So we reach the value of 1.19$\pm$0.11~Gyr for the white
dwarf cooling time of Procyon B.

On the other hand, the $Z$ = 0 envelope sequence with otherwise
benchmark assumptions yields a considerably larger age of 1.391~Gyr.
Since this is inconsistent with ultraviolet observations of carbon and
heavy elements in the ultraviolet spectrum, we consider a metal-free
envelope an unrealistic assumption.  Note that the atmospheric values
determined by Provencal et al. (2002) already have a heavy element
abundance (dominated by carbon) of Z$\sim$0.00001.  This parameter
increases greatly below the convection zone throughout the
partially-degenerate helium envelope (following profile of carbon
dredge-up), until it reaches the C/O core.

Likewise, we do not consider the assumption of a pure carbon core
realistic.  The combination of a pure carbon core with zero
metallicity envelope yields an even larger cooling age of 1.557~Gyr.
The cooling age listed for the alternative mass of 0.602 $M_{\odot}$,
assuming benchmark values of the other parameters above, is 1.406~Gyr.
A mass this large for the secondary disagrees by more than 2$\sigma$
from that derived from HST astrometry and we no longer consider it in
the rest of this paper. These several possibilities in this paragraph
are not factored into the error bars in the preceding paragraph.

The difference between the age of the Procyon system and the cooling
time of the white dwarf component, 0.68$\pm$0.17 Gyr, corresponds to
the sum of the pre-main sequence lifetime and the main sequence
lifetime (plus the much shorter red giant phases) of the Procyon B
progenitor.  Using again our TYCHO evolutionary code, we thus find
that the initial mass ($M_i$) of the secondary is
2.59$_{-0.18}^{+0.22}$ $M_{\odot}$, if the envelope Z value is fixed
at 0.001.  However, if one allows for the extreme upper limit to the
cooling time from the assumption of Z(env) = 0, the error bars for the
progenitor mass increase to $_{-0.26}^{+0.44}$ $M_{\odot}$. The quoted
errors on $M_i$ are the quadrature sums of $-$0.21,+0.40 due to the
incertainty in the cooling time of the white dwarf, and $-$0.16,+0.20
due to systematic error in the age of the binary system (Table~2).  In
Figure~2, discussed in detail in the next section, the solid
horizontal error bars for Procyon~B are for the case of Z(env) fixed
at 0.001, the extended light error bars reflect the increase in
cooling time for the limiting case Z(env) = 0.

\subsection{Remarks on the dependence of the cooling rate on the 
envelope opacities}

It may be worthwhile at this point to attract the attention of the
reader on some subtleties of white dwarf cooling which leave their
signature on the data presented in Table 1. For instance, as is well
known (and revealed again in the table), pure C core models
systematically cool slower than models having cores made of heavier
elements, all other things being the same. That a pure carbon core takes
longer to cool is well understood, because the thermal energy in a 
given mass is locked up in more nucleons than for pure oxygen or a
mixture. However, that the $Z$ = 0 envelope takes longer to cool than one
with heavy elements is not so intuitive, because one might think that
the more opaque envelope with heavy elements would let the interior
energy out more slowly.  Our calculations show that cooling proceeds
in two phases.  At first the less opaque model does let the thermal
energy escape more quickly, and core temperature decreases more rapidly.
Despite this, however, the decrease in luminosity has to be slower for
the less opaque model because of that extra energy that is made
available to fuel the luminosity. Hence, in a first phase, the less
opaque model actually cools {\it more slowly} to a given luminosity or
$T_{\rm eff}$. However then, after the less opaque star has let more
of its thermal energy escape, there is no turning back and the price
to pay, in a second phase, is that it will now cool {\it faster} the
rest of the way than its more opaque counterpart.  In the case of
Procyon B at $T_{\rm eff}$ = 7,740~K, it is still warm emough to be in
the first phase of that ``relative'' cooling and, hence, this is why
the cooling age is larger for the less opaque model.  If the star had
a $T_{\rm eff}$ of $\sim$5,000~K, then indeed the cooling age based on
the less opaque envelope sequence would be shorter than than of the
more opaque, as the conventional wisdom would expect.  These arguments
may be discerned from a careful perusal of Section III(a), subsection
(ii) of Tassoul, Fontaine and Winget (1990), although this discussion
compares an opaque envelope of hydrogen with less opaque counterparts.

\section{Procyon~B and the Initial to Final Mass Ratio (IFMR) }

As discussed in the last paragraph of the Introduction, the
initial-to-final mass relation (IFMR) for progenitors and white dwarfs
is fundamental in understanding a stellar population.  This study of
Procyon adds a data point near the low mass end with well determined
stellar parameters; it may be valuable for testing whether differences
exist in the IFMR for the distributions of hydrogen and helium-rich
atmosphere white dwarfs.

Table~2 lists final ($M_f$) and initial ($M_i$) masses in $M_{\odot}$,
cooling times (log in years), and estimated errors, first for
Procyon~B and Sirius~B. The remaining data points come from white
dwarfs found in Galactic star clusters, in order of increasing age
(see footnotes).  The collective data set is shown in Figure~2.
Again, the age of the population is obtained generally from fitting
the main sequence turnoff of the cluster color-magnitude diagram.

As given in Ferrario et al. (2005), the errors in $M_i$ fall into two
categories -- observational (obs) or random, and systematic (sys), as
listed in Table~2.  The observational errors devolve from
uncertainties in the $M_f$ from fitting the Balmer lines (or, in one
case, He~I lines); the systematic errors come from uncertainties in
the cluster or binary system age.  Since the nuclear lifetime of the
progenitor is the difference between the systemic age and the white
dwarf cooling time, $M_i$ depends on the age as well.  Note that the
uncertainties in the age determinations for the young clusters lead to
large error bars in the estimates of the initial masses $M_i$.

We add GD~50 to the Pleiades ``moving group'', since Dobbie et
al. (2006) make a strong case based on astrometric and spectroscopic
data that this ultramassive white dwarf is associated with the star
formation event that created the Pleiades cluster.  They argue that it
evolved as a single star from a progenitor of 6.3 $M_{\odot}$, and that
this may represent the first observational evidence that single-star
evolution can produce white dwarfs with, in this case, a mass of
1.27 $M_{\odot}$ (Bergeron, Saffer, \& Liebert 1992) or
1.2(+0.07,$-$0.08) $M_{\odot}$ (Bergeron et al. 1991).  They make a
weaker case that the massive white dwarf PG~0136+251,
1.20$\pm$0.03 $M_{\odot}$ (Bergeron, Saffer, \& Liebert 1992), may also
be related to the Pleiades. However, information is lacking on the
total ($UVW$) space motion of this object, so we do not include it
here.  We adopt an age of 125$\pm$25~Myr for the Pleiades, which
depends on the treatment of the convective core.  This excludes a
systematic uncertainty to the age which will be true for all clusters
with main sequence turnoff stars of mass above $\sim$~1.3 $M_{\odot}$.

Williams, Bolte, \& Koester (2004, 2009) present studies of white
dwarfs in the cluster NGC~2168~(M35); this work supercedes the data
listed in Ferrario et al. (2005).  In the updated listing, 2168-22 is
excluded since it is likely magnetic, and the weak Balmer lines could
not be fit by the authors.  The age of the cluster is believed to be
175$\pm$50~Myr (Sung and Bessell 1999, von Hippel 2005).  

For NGC~3532 (age 300$\pm$25~Gyr) we take the view that the analysis
with one of the 8.2~m VLT telescopes by Dobbie et al. (2009b)
supercedes the earlier ESO observations of Koester \& Reimers (1993).
For NGC~2099 (age 490$\pm$70~Myr), we note that star 2099-14 at
0.45$\pm$0.08 $M_{\odot}$ likely has a helium core, implying
formation from binary star evolution.  A new white dwarf in the
500$\pm$100~Myr Coma Berenices open cluster (Melotte 111), called
1216+260, has been studied by Dobbie et al. (2009a).  Based on
assuming a CO core and a thick H layer for this DA, they calculate the
parameters listed in Table~2. 

For the Praesepe and Hyades clusters, likely members of the Hyades
``moving group'' (Eggen 1958, Zuckerman \& Song 2004), we adopt as the
age 625$\pm$50~Myr.  The white dwarfs from the Hyades were studied by
Claver et al. (2001) and Ferrario et al. (2005).  The only DB star
among the cluster white dwarfs used here is 0437+138 in the Hyades.
The Praesepe cluster stars were studied by Claver et al. (2001),
Ferrario et al. (2005), Dobbie et al. (2004), and Casewell et
al. (2009).  For objects in common between these last two papers with
many common authors, we adopt the values of the last paper, since the
results were obtained with the UVES spectrograph on the VLT~UT2.  We
omit 0837+218 since Casewell et al. (2009) make a good case that this
is not a cluster member.

White dwarfs in older clusters overlapping and exceeding the likely
age of Procyon have been analyzed and published since the Ferrario et
al. work.  For the older clusters (as well as Procyon~B and Sirius~B),
the errors for $M_i$ are relatively smaller.

Kalirai et al. (2008) identified four likely members of the open
cluster NGC~7789, with age 1.40$\pm$0.14~Gyr, and found two likely
member white dwarfs in NGC~6819 with age 2.50$\pm$0.25~Gyr.  We note
that the signal-to-noise ratio of the spectra with fitted Balmer lines
are low compared to P. Bergeron's usual standards; we quote the listed
errors here.  Kalirai et al. (2007) analyzed a number of the brightest
white dwarfs in the very old Galactic cluster NGC~6791
(8.5$\pm$1.0~Gyr); however, only 6791-7 was the only likely member not
having a very low mass determination and likely helium core.  The
authors note that enhanced mass loss in the red giant phase, which is
the likely origin of the helium white dwarfs, probably also occurred
in this object, despite the fact that it retained a hydrogen
atmosphere.

Finally, Kalirai et al. (2009) analyze a number of white dwarfs at the
tip of the observed sequence in the globular cluster Messier~4.  This
is of course a Population~II system with [Fe/H] = $-$1.10$\pm$0.01
(Mucciarelli et al. 2011), and likely age of 12.7$\pm$0.7~Gyr (Hansen
et al. 2002).  The basic conclusion is that the $\sim$0.8$\pm$0.05
$M_{\odot}$ stars now producing white dwarfs which have evolved to
approximately 0.53$\pm$0.01 $M_{\odot}$ (but we list the individual
determination for each of the latter).

Procyon B has a final mass below nearly all others of similar initial
mass.  It is one of only two white dwarfs here with a helium
atmosphere.  The other is the DB in the Hyades mentioned earlier,
which has a mass similar to the other DA stars in that cluster and the
Praesepe.  The dashed line plots the simple linear relation from
Ferrario et al. (2005) based on a fit to white dwarfs in more massive
clusters available at that time.  The Procyon~B point lies more than
0.1 $M_{\odot}$ or several $\sigma$ below this line in the sense of
having a lower final mass.

We should note that the close, cool double-DC binary G~107-70AB
(Harrington, Christy, \& Strand 1981) appears to be a similar
situation at first glance.  It was barely resolved from ground-based
imaging at the U.S. Naval Observatory, Flagstaff Station.  It is no
surprise that this old result is superceded by 13 years of HST imaging
(1995.8-2008.8), as of the latest fit -- Schaefer et al. (2006, and
private communication).  The preliminary period is 18.546$\pm$0.082
years, $a$ = 0.666" , and $M_{tot}$ = 1.191$\pm$0.057 $M_{\odot}$
(using the Hipparcos parallax of 0.0896$\pm$0.0014").  Since only a
small difference in magnitude between the two components is indicated,
this suggests two components each of not quite 0.6 $M_{\odot}$.

Using the integrated light from the two stars, Bergeron, Ruiz, \&
Leggett (1997) estimate 4900~K, log $g$  = 7.35, and their broad-band
$BVRIJHK$ energy distribution fits an H-rich atmosphere.  That is, the
spectrum would have been type DA if the temperatures of the two stars
were warmer.  The HST astrometry above suggests two white dwarfs each
of mass near the well known peak of the DA mass distribution (Koester,
Schulz, \& Weidemann 1979).  Note that the log $g$ value inferred
above is incorrect, since the method used by Bergeron et al. (1997) to
estimate a surface gravity assumes a single star.  Thus G~107-70AB are
not cooler analogs of Procyon~B.

We have no explanation for why the Procyon~B remnant appears so
undermassive vs the initial mass we have derived.  Suppose that
virtually all of the helium envelope of the Procyon~B nuclear
progenitor could have been been lost in the event (late helium-shell
flash?) that also removed all of the hydrogen. However, stellar models
of the asymptotic giant branch phase show that the maximum helium
envelope mass remaining would have been only $\sim$0.01$M_{\odot}$ if
left intact.

Thus the loss of the entire helium envelope fails to account for the
total mass shortfall by an order of magnitude.  Rather, the evolution
of the (presumably) carbon-oxygen core was evidently truncated before
it could build up to a more normal mass for a single star with the
indicated initial mass.  The dilemma suggests that close binary
evolution may have been involved, but the existing binary consists of
two stars in an elliptical orbit of long period.

Hence, we are left with an interesting problem in (binary) stellar
evolution.  Note that there are still rather few stars analyzed from
Population~I clusters of age older than or similar to Procyon.  In the
future the numerous white dwarfs in the old disk clusters M~67 and
NGC~188 will make useful additions to round out the low mass end of
such a diagram.

\acknowledgments

We gratefully acknowledge Gail Schaefer of the CHARA array at Georgia
State University, and Howard E. Bond of STScI, for allowing us to
quote preliminary HST results which fix the masses of Procyon A and B,
and for results on G~107-70AB.  This work was originally supported by
the National Science Foundation through grant AST-0307321 (JL and
KAW). GF wishes to acknowledge the contribution of the Canada Research
Chair Program.

\clearpage 

\begin{deluxetable}{cccccccc}
\tablenum{1}
\tabletypesize{\footnotesize}
\tablecaption{Parameters of Cooling Models at $T_{\rm eff}$ = 7,940~K, 7,740~K,
and 7,540~K}  
\tablewidth{0pt}
\tablehead{
\colhead{ID} & 
\colhead{Core} & 
\colhead{log $M_{\rm He}/M$} & 
\colhead{$$Z$$} & 
\colhead{$M/M_{\odot}$} & 
\colhead{log $g$} & 
\colhead{$R/R_{\odot}$} & 
\colhead{Age (Gyr)}
}
\startdata 
  &     &        &       &       & 7.9676 & 0.012777 & 1.1069 \\
1 & C/O & $-$2.5 & 0.001 & 0.553 & 7.9681 & 0.012768 & 1.1870 \\
  &     &        &       &       & 7.9687 & 0.012760 & 1.2772 \\
  &     &        &       &       & 7.9297 & 0.013077 & 1.0516 \\
2 & C/O & $-$2.5 & 0.001 & 0.531 & 7.9303 & 0.013068 & 1.1274 \\
  &     &        &       &       & 7.9309 & 0.013059 & 1.2122 \\
  &     &        &       &       & 8.0045 & 0.012485 & 1.1650 \\
3 & C/O & $-$2.5 & 0.001 & 0.575 & 8.0050 & 0.012479 & 1.2501 \\
  &     &        &       &       & 8.0055 & 0.012471 & 1.3438 \\
  &     &        &       &       & 7.9637 & 0.012834 & 1.2336 \\
4 & C   & $-$2.5 & 0.001 & 0.553 & 7.9642 & 0.012826 & 1.3236 \\
  &     &        &       &       & 7.9648 & 0.012817 & 1.4251 \\
  &     &        &       &       & 7.9709 & 0.012728 & 1.0992 \\
5 & O   & $-$2.5 & 0.001 & 0.553 & 7.9714 & 0.012720 & 1.1770 \\
  &     &        &       &       & 7.9720 & 0.012712 & 1.2650 \\
  &     &        &       &       & 7.9730 & 0.012696 & 1.2962 \\
6 & C/O & $-$2.5 & 0.0   & 0.553 & 7.9736 & 0.012688 & 1.3914 \\
  &     &        &       &       & 7.9742 & 0.012680 & 1.4990 \\
  &     &        &       &       & 7.9693 & 0.012752 & 1.4500 \\
7 & C   & $-$2.5 & 0.0   & 0.553 & 7.9699 & 0.012743 & 1.5573 \\
  &     &        &       &       & 7.9705 & 0.012734 & 1.6774 \\
  &     &        &       &       & 7.9762 & 0.012651 & 1.2634 \\
8 & O   & $-$2.5 & 0.0   & 0.553 & 7.9767 & 0.012643 & 1.3544 \\
  &     &        &       &       & 7.9772 & 0.012635 & 1.4565 \\
  &     &        &       &       & 7.9661 & 0.012798 & 1.0705 \\
9 & C/O & $-$2.0 & 0.001 & 0.553 & 7.9667 & 0.012790 & 1.1488 \\
  &     &        &       &       & 7.9673 & 0.012781 & 1.2374 \\
  &     &        &       &       & 7.9687 & 0.012756 & 1.1521 \\
10& C/O & $-$3.0 & 0.001 & 0.553 & 7.9692 & 0.012747 & 1.2347 \\
  &     &        &       &       & 7.9698 & 0.012739 & 1.3266 \\
  &     &        &       &       & 8.0485 & 0.012145 & 1.3120 \\
11& C/O & $-$2.5 & 0.001 & 0.602 & 8.0489 & 0.012138 & 1.4063 \\
  &     &        &       &       & 8.0494 & 0.012132 & 1.5112 
\enddata 
\end{deluxetable}

\clearpage

\begin{deluxetable}{llllllllr}
\tablenum{2}
\tablecaption{Final -- Initial Mass ($M_f$/$M_i$) Determinations } 
\tablewidth{0pt}
\tablehead{
\colhead{Star Name} & \colhead{$M_f$} & \colhead{d$M_f$} &
\colhead{log~$\tau$} & \colhead{dlog$\tau$} & 
\colhead{$M_i$} & \colhead{$\pm$~d$M_i(obs)$} &
\colhead{$\pm$~d$M_i(sys)$} & Ref 
}

\startdata 

Procyon~B & 0.553 & 0.022 & 9.075 & 0.074 & 2.59 & 0.21,0.40 & 
0.16,0.20 & 1 \\
Sirius~B & 1.000 & 0.020 & 8.091 & 0.020 & 5.056 & 0.171,0.262 & 
0.213,0.273 & 2 \\
LB~1497 & 1.023 & 0.026 & 7.781 & 0.062 & 6.542 & 0.346,0.458 & 
0.866,1.614 & 3 \\ 
GD~50 & 1.264 & 0.017 & 7.785 & 0.045 & 6.3 & 2.5,0.8 & 
--,-- & 4,5 \\  
NGC~2516-1 & 0.931 & 0.098 & 7.760 & 0.230 & 5.411 & 1.017,0.500 & 
0.524,0.386 & 6 \\ 
NGC~2516-2 & 1.004 & 0.058 & 7.621 & 0.179 & 5.096 & 0.406,0.239 & 
0.415,0.311 & 6 \\ 
NGC~2516-3 & 0.969 & 0.039 & 7.922 & 0.079 & 6.141 & 0.692,0.427 & 
0.924,0.587 & 6 \\ 
NGC~2516-5 & 1.054 & 0.063 & 7.883 & 0.136 & 5.902 & 0.986,0.524 & 
0.786,0.514 & 6 \\ 
NGC~2168-1 & 0.873 & 0.091 & 7.228 & 0.260 & 4.39 & 0.23,0.09 & 
0.35,0.27 & 7 \\ 
NGC~2168-2 & 1.015 & 0.067 & 7.657 & 0.202 & 4.79 & 0.47,0.26 & 
0.46,0.36 & 7 \\ 
NGC~2168-5 & 0.916 & 0.075 & 6.103 & 0.088 & 4.21 & 0.00,0.00 & 
0.29,0.23 & 7 \\ 
NGC~2168-6 & 0.877 & 0.096 & 6.077 & 0.082 & 4.21 & 0.00,0.00 & 
0.29,0.23 & 7 \\
NGC~2168-11 & 0.802 & 0.096 & 8.047 & 0.173 & 6.63 & 0.00,1.38 & 
1.89,0.96 & 7 \\ 
NGC~2168-12 & 0.922 & 0.092 & 7.314 & 0.281 & 4.43 & 0.30,0.13 & 
0.36,0.27 & 7 \\ 
NGC~2168-14 & 1.010 & 0.072 & 7.865 & 0.167 & 5.32 & 0.95,0.44 &
0.73,0.48 & 7 \\ 
NGC~2168-15 & 0.888 & 0.088 & 7.551 & 0.236 & 4.63 & 0.38,0.22 & 
0.41,0.32 & 7 \\ 
NGC~2168-27 & 1.022 & 0.072 & 7.840 & 0.170 & 5.22 & 0.71,0.42 & 
0.68,0.45 & 7 \\ 
NGC~2168-29 & 0.882 & 0.078 & 7.331 & 0.235 & 4.44 & 0.23,0.12 &
0.36,0.28 & 7 \\
NGC~2168-30 & 1.011 & 0.122 & 7.808 & 0.289 & 5.12 & 1.49,0.55 & 
0.63,0.43 & 7 \\
NGC~2287-2 & 0.910 & 0.040 & 7.908 & 0.075 & 4.450 & 0.580,0.380 & 
--,-- & 8 \\
NGC~2287-5 & 0.910 & 0.040 & 7.964 & 0.071 & 4.570 & 0.640,0.430 & 
--,-- & 8 \\ 
NGC~3532-1 & 0.86  & 0.04 & 7.778 & 0.095 & 3.830 & 0.18,0.15 & 
--,-- & 9 \\
NGC~3532-5 & 0.820 & 0.04 & 7.580 & 0.115 & 3.71 & 0.150,0.130 & 
--,-- & 9 \\
NGC~3532-9 & 0.760 & 0.04 & 7.00 & 0.113 & 3.57 & 0.12,0.11 & 
--,-- & 9 \\ 
NGC~3532-10 & 0.96 & 0.04 & 8.173 & 0.060 & 4.58 & 0.47,0.33 & 
--,-- & 9 \\ 
NGC~2099-2 & 0.69  & 0.10 & 7.961 & 0.165 & 2.76 & 0.08,0.05 & 
0.162,0.201 & 10 \\ 
NGC~2099-3 & 0.76 & 0.13 & 8.190 & 0.190 & 2.92 & 0.16,0.14 & 
0.197,0.258 & 10 \\ 
NGC~2099-4 & 0.87 & 0.15 & 8.428 & 0.217 & 3.02 & 0.26,0.19 & 
0.324,0.495 & 10 \\ 
NGC~2099-5 & 0.83 & 0.14 & 8.271 & 0.188 & 3.02 & 0.26,0.19 & 
0.219,0.305 & 10 \\ 
NGC~2099-7 & 0.88 & 0.19 & 8.378 & 0.289 & 3.26 & 0.78,0.40 & 
0.278,0.405 & 10 \\ 
NGC~2099-9 & 0.61 & 0.05 & 8.268 & 0.067 & 2.97 & 0.06,0.07 & 
0.218,0.302 & 10 \\ 
NGC~2099-10 & 0.74 & 0.04 & 8.090 & 0.068 & 2.81 & 0.04,0.02 & 
0.178,0.225 & 10 \\ 
NGC~2099-11 & 0.96 & 0.06 & 8.149 & 0.095 & 2.86 & 0.07,0.07 & 
0.188,0.242 & 10 \\ 
NGC~2099-12 & 0.55 & 0.07 & 8.394 & 0.126 & 3.30 & 0.30,0.22 & 
0.290,0.429 & 10 \\ 
NGC~2099-13 & 0.79 & 0.05 & 8.229 & 0.074 & 2.93 & 0.06,0.07 & 
0.207,0.278 & 10 \\ 
NGC~2099-14 & 0.45 & 0.08 & 8.487 & 0.085 & 3.31 & 0.20,0.26 & 
0.400,0.674 & 10 \\
NGC~2099-16 & 0.83 & 0.06 & 8.689 & 0.082 & 5.20 & 7.0,1.20 & 
3.737,-- & 10 \\ 
1216+260 & 0.90 & 0.04 & 8.560 & 0.052 & 4.77 & 0.97,5.37 & 
--,-- & 11 \\ 
0352+098 & 0.719 & 0.030 & 8.270 & 0.046 & 3.094 & 0.045,0.052 & 
0.114,0.134 & 12 \\
0406+169 & 0.806 & 0.031 & 8.488 & 0.046 & 3.465 & 0.109,0.147 & 
0.172,0.225 & 12 \\   
0421+162 & 0.680 & 0.031 & 7.970 & 0.055 & 2.892 & 0.020,0.023 & 
0.090,0.103 & 12 \\
0425+168 & 0.705 & 0.031 & 7.549 & 0.077 & 2.789 & 0.011,0.010 & 
0.079,0.088 & 12 \\
0431+125 & 0.652 & 0.032 & 7.752 & 0.068 & 2.825 & 0.014,0.017 & 
0.083,0.093 & 12 \\
0437+138 & 0.740 & 0.060 & 8.470 & 0.047 & 3.663 & 0.030,0.030 &  
--,-- & 13 \\ 
0438+108 & 0.684 & 0.031 & 7.203 & 0.060 & 2.757 & 0.003,0.004 & 
0.076,0.084 & 12 \\ 
0833+194 & 0.721 & 0.043 & 8.431 & 0.049 & 3.32 & 0.33,0.22 & 
--,-- & 14 \\ 
0836+197 & 0.909 & 0.030 & 8.136 & 0.049 & 2.981 & 0.031,0.035 & 
0.101,0.115 & 15 \\ 
0836+199 & 0.831 & 0.037 & 8.612 & 0.069 & 3.997 & 0.350,0.652 & 
0.297,0.428 & 15 \\ 
0836+199 & 0.752 & 0.044 & 8.489 & 0.049 & 3.46 & 0.43,0.27 & 
--,-- & 14 \\
0836+201 & 0.620 & 0.031 & 8.154 & 0.050 & 2.993 & 0.028,0.046 & 
0.102,0.117 & 15 \\ 
0837+185 & 0.804 & 0.044 & 8.504 & 0.050 & 3.50 & 0.48,0.29 & 
--,-- & 14 \\ 
0837+199 & 0.819 & 0.032 & 8.351 & 0.048 & 3.194 & 0.081,0.062 & 
0.153,0.129 & 15 \\ 
0837+199 & 0.737 & 0.043 & 8.253 & 0.052 & 3.07 & 0.20,0.15 & 
--,-- & 14 \\ 
0840+190 & 0.849 & 0.045 & 8.566 & 0.050 & 3.73 & 0.71,0.39 & 
--,-- & 14 \\ 
0840+200 & 0.761 & 0.033 & 8.522 & 0.045 & 3.572 & 0.186,0.129 & 
0.255,0.196 & 15 \\ 
0840+200 & 0.721 & 0.043 & 8.420 & 0.048 & 3.30 & 0.32,0.21 & 
-- & 14 \\ 
0843+184 & 0.823 & 0.045 & 8.530 & 0.051 & 3.59 & 0.55,0.33 &
--,-- & 14 \\
NGC~7789-4 & 0.560 & 0.020 & 8.061 & 0.032 & 2.080 & 0.080,0.080 & 
--,-- & 16 \\ 
NGC~7789-5 & 0.600 & 0.030 & 6.903 & 0.055 & 2.020 & 0.070,0.140 & 
--,-- & 16 \\ 
NGC~7789-6 & 0.720 & 0.030 & 8.204 & 0.045 & 2.100 & 0.090,0.090 & 
--,-- & 16 \\ 
NGC~7789-8 & 0.640 & 0.040 & 7.462 & 0.080 & 2.020 & 0.090,0.110 & 
--,-- & 16 \\ 
NGC~6819-6 & 0.530 & 0.020 & 7.591 & 0.035 & 1.600 & 0.060,0.050 & 
--,-- & 17 \\ 
NGC~6819-7 & 0.560 & 0.020 & 8.155 & 0.034 & 1.620 & 0.070,0.050 & 
--,-- & 17 \\ 
NGC~6791-7 & 0.530 & 0.020 & 8.176 & 0.060 & 1.160 & 0.040,0.030 & 
--,-- & 18 \\ 
M~4-0 & 0.520 & 0.040 & 7.602 & 0.070 & 0.800 & 0.050,0.050 & 
--,-- & 19 \\ 
M~4-4 & 0.500 & 0.030 & 7.279 & 0.024 & 0.800 & 0.050,0.050 & 
--,-- & 19 \\ 
M~4-6 & 0.590 & 0.040 & 7.255 & 0.050 & 0.800 & 0.050,0.050 & 
--,-- & 19 \\ 
M~4-15 & 0.550 & 0.040 & 7.322 & 0.043 & 0.800 & 0.050,0.050 & 
--,-- & 19 \\ 
M~4-20 & 0.510 & 0.050 & 7.690 & 0.075 & 0.800 & 0.050,0.050 & 
--,-- & 19 \\ 
M~4-24 & 0.510 & 0.030 & 7.204 & 0.028 & 0.800 & 0.050,0.050 & 
--,-- & 19 \\  

\enddata 
\tablecomments{
 (1) Procyon, 1.87~Gyr, this paper;  
 (2) Sirius, 237.5$\pm$12.5, Liebert et al. (2005); 
 (3) The Pleiades, 125$\pm$25~Myr, Claver et al. (2001);  
 (4) The Pleiades, 125$\pm$25~Myr, Bergeron et al. (2002); 
 (5) The Pleiades, 125$\pm$25~Myr, Dobbie et al. (2006);   
 (6) NGC~2516, 158$\pm$20~Myr, Ferrario et al. (2005);  
 (7) NGC~2168, 175$\pm$25~Myr, Williams et al. (2004, 2009)
 (8) NGC~2287, 243$\pm$40~Myr, Dobbie et al. (2009b);  
 (9) NGC~3532, 300$\pm$25~Myr, Dobbie et al. (2009b);  
(10) NGC~2099, 490$\pm$70~Myr, Ferrario et al. (2005);  
(11) Coma Berenices (Melotte~111), 500$\pm$100~Myr, Dobbie et al. (2009a) 
(12) The Hyades, 625$\pm$50~Myr, Claver et al. (2005);  
(13) DB in the Hyades, 625$\pm$50~Myr, Bergeron et al. (2011);   
(14) The Praesepe, 625$\pm$50~Myr, Casewell et al. (2009) 
(15) The Praesepe, 625$\pm$50~Myr, Claver et al. (2001); 
(16) NGC~7789, 1.4$\pm$0.14~Gyr, Kalirai et al. (2008);  
(17) NGC~6819, 2.5$\pm$0.25~Gyr, Kalirai et al. (2008);  
(18) NGC~6791, 8.5$\pm$1~Gyr, Kalirai et al. (2007);  
(19) Messier~4, $\sim$12.7~Gyr, Kalirai et al. (2009); } 

\end{deluxetable}

\clearpage

\centerline{\bf{FIGURE CAPTIONS}}

\noindent Fig. 1 --- TYCHO evolutionary tracks in a log luminosity
vs. log radius diagram beginning including the pre-main sequence at
the Asplund, Grevesse, \& Sauval (2005, 2009 =A09) primordial solar
abundance of Z=0.0122 for masses of (in descending order of
luminosity) 1.499 $M_{\odot}$ (black), the favored preliminary HST
astrometric value, 1.48 $M_{\odot}$ (red), the best fit to the observed
luminosity, 1.463 $M_{\odot}$ (green), and 1.42 $M_{\odot}$ (blue).

\noindent Fig. 2 --- The initial to final mass relation (IFMR) for
Procyon~B, Sirius~B and the white dwarfs in a number of star clusters
discussed in the text.  Procyon~B and Sirius~B (from Paper~I) are
plotted as black, filled circles. The total data set for older open
clusters from Kalirai et al. (2007, 2008, 2009) are also plotted with
filled symbols -- NGC~7789 (red), NGC~6819 (green), NGC~6791 (blue)
and Messier~4 (magenta).  Shown with open circles -- taken from
Ferrario et al.  (2005) and Paper~I -- are the Pleiades (red),
NGC~2516 (cyan), NGC~2168 (magenta), the Hyades (blue) and Praesepe
(cyan), NGC~3532 (green), and NGC~2099 (yellow).  The error bars
include the best estimates of the cluster ages.  The magenta line
plots the simple linear relation from Ferrario et al. (2005).  The
Procyon~B remnant mass appears low relative to the others, as discussed
in the text.

\clearpage 
\begin{figure}[p]
\plotone{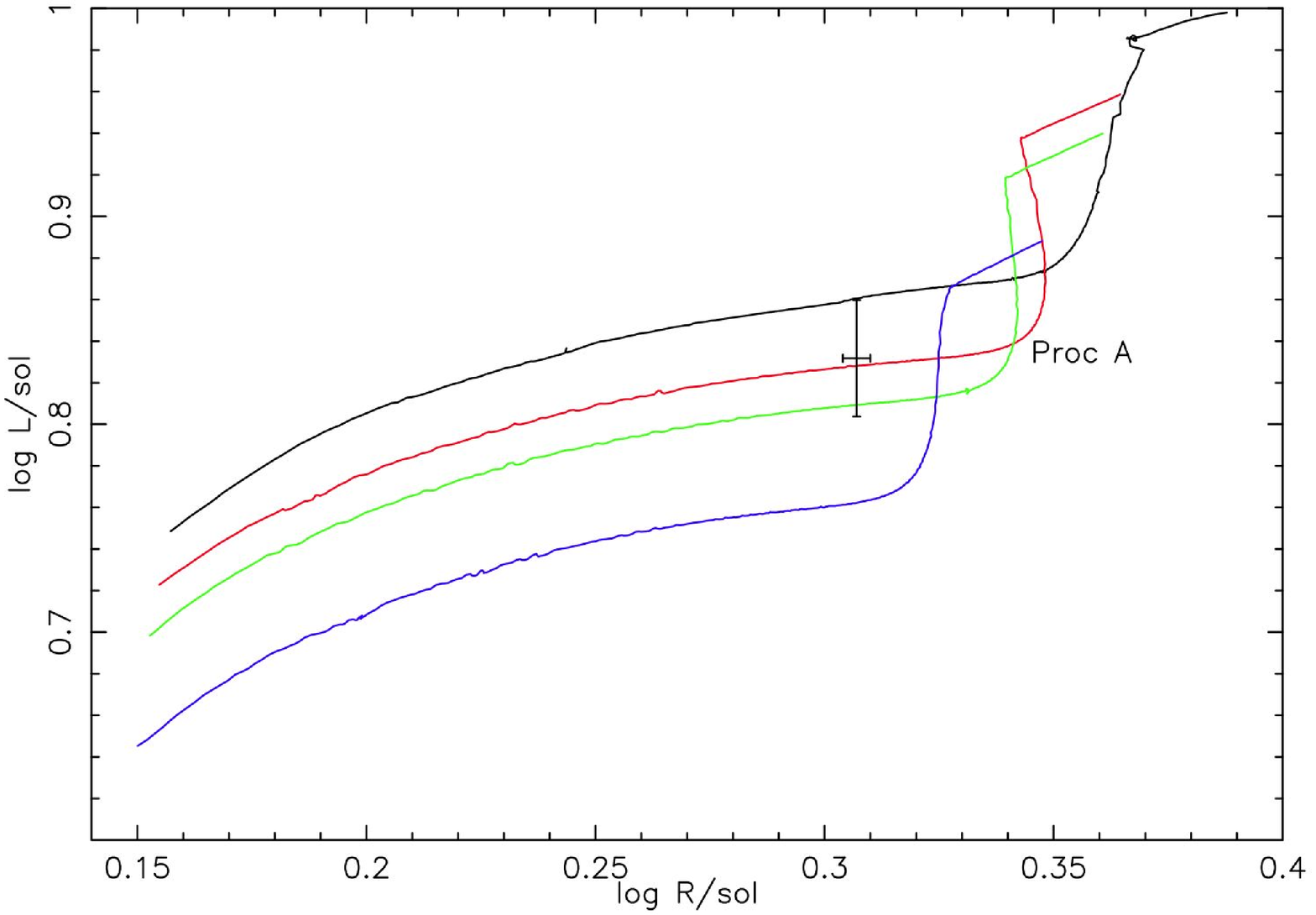}
\begin{flushright}
Figure 1
\end{flushright}
\end{figure}

\clearpage
\begin{figure}[p]
\plotone{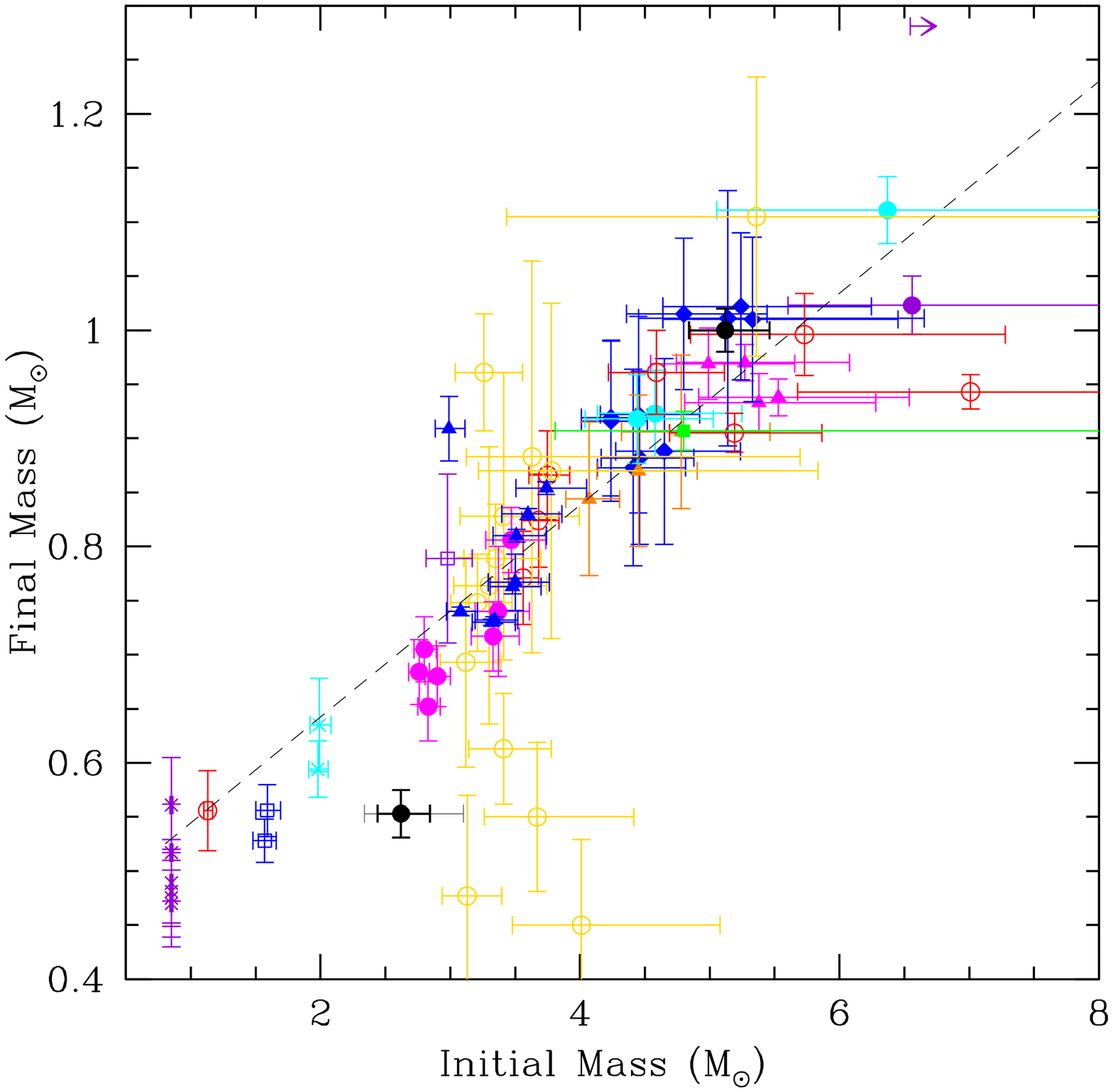}
\begin{flushright}
Figure 2
\end{flushright}
\end{figure}

\end{document}